\documentclass[review]{elsarticle}

\usepackage{subfigure}
\usepackage{graphicx}
\usepackage{amsmath}
\usepackage{amssymb}
\usepackage{float}
\usepackage{soul} 

\usepackage{lineno}
\usepackage{multirow}
\usepackage{enumitem}
\usepackage{natbib}

\begin{document}

\begin{frontmatter}

\title{asKAN: Active Subspace embedded Kolmogorov-Arnold Network}

\author{Zhiteng Zhou\footnote{Postdoctoral research scientist, LNM, Institute of Mechanics, Chinese Academy of Sciences.}, Zhaoyue Xu\footnote{Postdoctoral research scientist, LNM, Institute of Mechanics, Chinese Academy of Sciences.},Yi Liu\footnote{Postdoctoral research scientist, LNM, Institute of Mechanics, Chinese Academy of Sciences.}, Shizhao Wang\footnote{Professor, LNM, Institute of Mechanics, Chinese Academy of Sciences, corresponding author, wangsz@lnm.imech.ac.cn.}}
\address{LNM, Institute of Mechanics, Chinese Academy of Sciences, Beijing,China, 100190}
\address{School of Engineering Sciences, University of Chinese Academy of Sciences, Beijing,China, 100049}

\begin{abstract}
The Kolmogorov-Arnold Network (KAN) has emerged as a promising neural network architecture for small-scale AI+Science applications. However, it suffers from inflexibility in modeling ridge functions, which is widely used in representing the relationships in physical systems. This study investigates this inflexibility through the lens of the Kolmogorov-Arnold theorem, which starts the representation of multivariate functions from constructing the univariate components rather than combining the independent variables. Our analysis reveals that incorporating linear combinations of independent variables can substantially simplify the network architecture in representing the ridge functions. Inspired by this finding, we propose active subspace embedded KAN (asKAN), a hierarchical framework that synergizes KAN's function representation with active subspace methodology. The architecture strategically embeds active subspace detection between KANs, where the active subspace method is used to identify the primary ridge directions and the independent variables are adaptively projected onto these critical dimensions. The proposed asKAN is implemented in an iterative way without increasing the number of neurons in the original KAN. The proposed method is validated through function fitting, solving the Poisson equation, and reconstructing sound field. Compared with KAN, asKAN significantly reduces the error using the same network architecture. The results suggest that asKAN enhances the capability of KAN in fitting and solving equations in the form of ridge functions.

	
\end{abstract}

\begin{keyword}
Active Subspace Method; Kolmogorov-Arnold Network; Sound reconstruction; Intrinsically Low-Dimensional Problems
\end{keyword}

\end{frontmatter}

\section{Introduction}





Kolmogorov-Arnold network~\cite{liu2024kan}(KAN) is a promising neural network architecture alternative to Multi-Layer Perceptron (MLP).  
KANs employ fully connected structures consisting of \emph{nodes} (``neurons'') connected through \emph{edges} (``weights''). 
Each \emph{node} processes the data flows by passing the weighted sum of the inputs through an activation function.
Unlike traditional MLP, KANs replace fixed activation functions on \emph{nodes} with learnable activation functions on \emph{edges}. 
Replacement of the activation function not only endows the KAN network with improved accuracy and interpretability on small-scale AI + Science tasks, but also reduces the overall scale of the network~\cite{liu2024kan,hou2024comprehensive}.

The advantage of KAN has been reported across different computational tasks. In fitting high-frequency oscillatory functions, KANs achieve  comparable or better precision than MLPs, while using only half the parameters~\cite{liu2024kan1}. 
This improvement in function fitting has motivated extensive applications to increase accuracy, handle sparse data, and enhance efficiency. 
For example, Mostajeran and Faroughi~\cite{mostajeran2024epi} modeled stainless steel deformation under cyclic loads using Chebyshev-KAN, achieving a correlation of 0.99 with experimental stress-strain curves (versus 0.92 for classical models).
Zhou et al.~\cite{zhou2025kolmogorov} applied KAN to model wall pressure fluctuations. Compared to MLPs, KAN demonstrated superior capability in accurately reconstructing the wavenumber-frequency spectrum over zero-pressure gradient regions on a revolution from sparse data.
Bozorgasl and Chen~\cite{bozorgasl2024wav} improved bearing fault diagnosis by embedding wavelet functions in KAN, improving the classification accuracy from 85\% to 97\% while halving training time. 



KANs are constructed based on the Kolmogorov-Arnold representation theorem~\cite{kolmogorov1961representation}, which states that a multivariate continuous function on a bounded domain can be expressed as superpositions and compositions of a finite number of univariate functions.
Although the above work reports the advantages of KANs in function fitting, superpositions and composition of a finite number of univariate functions usually suffers from the lack of flexibility in representing functions approximated by ridge functions (see the details in Section 2.1).
The lack of flexibility origins from the feature of the KAN structure, since the first step of KAN is to compute activation function rather than the linear combination of the independent variables. 
This feature prevents KAN from directly identifying the relevant directions of ridge functions, which are critical in representing the relations in physical systems~\cite{constantine2015active,xu2022artificial,zhang2024clustering}. 


The aim of this work is to enhance the flexibility of KANs in representing multivariate functions which can be approximated by ridge functions. 
We propose to embed the active subspace method to KAN (hereinafter referred to as asKAN), where the active subspace method detects the primary directions of the ridge functions by evaluating the gradient. 
By projecting the initial independent variables onto these primary directions, we obtain a set of new independent variables that are fed back into KAN for training. 
We implement the asKAN in an iterative way, resulting in a compact low-dimensional neural network by identifying the primary directions in the multivariate function.
The remainder of this paper is organized as follows. The asKAN is introduced in Section 2. Validation of asKAN for function fitting and partial differential equation (PDE) solving are reported in Section 3. Finally, conclusions are drawn in Section 4.



\section{Active subspace embedded Kolmogorov-Arnold networks}
We first illustrate the  inflexibility of Kolmogorov-Arnold representation theorem in handling ridge functions in Section 2.1. Then, we briefly summarize the features of KANs and the active subspace method closely related to this work in Sections 2.2 and 2.3, respectively. Finally, we report the architecture and implementation of asKAN proposed by this work to circumvent the inflexibility of KANs. 

\subsection{Inflexibility of Kolmogorov-Arnold representation theorem}

The Kolmogorov-Arnold representation theorem states that a multivariate continuous function $f(\mathbf{x})$, with $\mathbf{x} = [x_1,x_2, \ldots, x_N]^T$, can be decomposed into superpositions and compositions of finite univariate functions \(\varphi_q\) and \(\phi_{q,m}\)~\cite{kolmogorov1961representation}. Here, $N$ is the number of independent variables. The subscripts \(m\) and $q$ index the independent variables and univariate functions, respectively. The decomposition can be expressed as
\begin{equation}
f(\mathbf{x}) = \sum_{q=1}^{2N+1} \varphi_q \left( \sum_{m=1}^N \phi_{q,m}(x_m) \right), \quad m = 1, 2, \ldots, N.
\end{equation}
The above theorem states the existence of the representation, but it does not provide an algorithm for constructing the univariate functions. 
Although different variants of Kolmogorov-Arnold theorem have been developed, most of them are expressed in the form of the limits or sums of some infinite series of functions, which are not suitable for practical computations~\cite{braun2009constructive}.
Sprecher~\cite{sprecher1996numerical,sprecher1997numerical} provided an practical algorithm for approximately constructing the univariate functions by introducing $\phi_{q,m}= \alpha_m \zeta \left( {{x_m} + qa} \right)$ with appropriate value $\alpha_m$, $a \in \mathbb{R}$. However, this algorithm still suffers from inflexibility in representing the ridge functions. For example, the function $g(\mathbf{x}) = \exp \left( \cos \left( 3\pi (x_1 + x_2) \right) \right)$ is approximated by an intricate and prolix expression as follows, 

\begin{align}
{g({\bf{x}}) \approx \sum\limits_{q = 1}^5 {{\varphi _q}} {\mathop{\rm o}\nolimits} \sum\limits_{m = 1}^2 {{\alpha _m}} \zeta \left( {{x_m} + qa} \right),}
\end{align}
where ${{\alpha _1} = 1,{\alpha_m} = \sum\limits_{r = 1}^\infty  {{\gamma ^{ - (m - 1)\beta (r)}}} {\rm{ \;for\; }}m > 1{\rm{\;and\; }}\beta (r) = \left( {{N^r} - 1} \right)/(N - 1)}$, respectively. The function $\zeta$ is defined pointwise on a dense subset of terminating rational numbers $d_k \in \mathbb{Q}$,
\begin{align}
\zeta \left( {{d_k}} \right) = \sum\limits_{r = 1}^k \left( i_r-(\gamma-2)\langle i_r \rangle \right) {2^{ - {\left\langle {{i_r}} \right\rangle  {\sum\limits_{s = 1}^{r - 1} {\left( {\left[ {{i_s}} \right] \cdots  \cdot \left[ {{i_{r - 1}}} \right]} \right)} } }}} {\gamma ^{ - \beta \left( {r - {\left\langle {{i_r}} \right\rangle ( {\sum\limits_{s = 1}^{r - 1} {\left( {\left[ {{i_s}} \right] \cdots  \cdot \left[ {{i_{r - 1}}} \right]} \right)} } )}} \right)}}.
\end{align}
Here, $\gamma \ge 2N + 2$ is an integer for constructing the terminating rational numbers. 
$\left\langle {{i_r}} \right\rangle$ and $\left[ {{i_r}} \right]$ are defined as 
$\left\langle {{i_1}} \right\rangle = 0$,  $[\left. {{i_1}} \right] = 0$ and
\begin{align}
\left\langle {{i_r}} \right\rangle = \left\{ {\begin{array}{*{20}{l}}
0&{{\rm{ when\;\;\;  }}{i_r} = 0,1, \ldots ,\gamma  - 2}\\
1&{{\rm{ when\;\;\; }}{i_r} = \gamma  - 1}
\end{array},} \right.
\end{align}
\begin{align}
\left[ {{i_r}} \right] = \left\{ {\begin{array}{*{20}{l}}
0&{{\rm{ when\;\;\; }}{i_r} = 0,1, \ldots ,\gamma  - 3}\\
1&{{\rm{ when\;\;\; }}{i_r} = \gamma  - 2,\gamma  - 1}
\end{array}.} \right.
\end{align}
The function $\varphi_q$ in Eq. (2) can be constructed in an iterative way starting from an arbitrary continuous function $\omega \left( {{\bf{d}}_{{k_r}}^q; {{{{y}}_q}} } \right)$. The first order approximation to $\varphi_q$ can be expressed as 
\begin{align}
\varphi_q\left( {{y_q}} \right) = \frac{1}{{m + 1}}\sum {{g}} \left( {{{\bf{d}}_{{k_r}}}} \right)\omega \left( {{\bf{d}}_{{k_r}}^q; {{{{y}}_q}} } \right).
\end{align}
More details of the representation can be found in the work of Braun and Griebel~\cite{braun2009constructive}, where the continuity and monotonicity of the function $\zeta$ are improved. 

The inflexibility origins from the feature that the inner layer of the Kolmo-gorov-Arnold representation theorem starts from the construction of univariate functions, while the ridge function varies with the combination of independent variables. 
An approach to circumvent this inflexibility is to combine the independent variables before we construct the univariate functions.  For the cases reported in the above example, the representation can be significantly simplified if we combine the independent variables $x_1 + x_2$ and denote it as $z$. 
The resulting simplified representation $f(z) = g(\mathbf{x})$ can be decomposed as follows
\begin{align}
f(z) = \exp \left( \cos \left( 3\pi z \right) \right) = \sum_{q=1}^{3} \varphi_q \left(\sum_{m=1}^{1} \phi_{q,m}(z) \right),
\end{align}
where 
\begin{align}
\phi_{q,1}(z)&=  \cos \left( 3\pi z \right), \\
\varphi_q\left( {{y_q}} \right) &= \left\{ {\begin{array}{*{20}{l}}
\exp \left( y_q \right)&{{\rm{ when\;\;\; }}{q} = 1},\\
0&{{\rm{ when\;\;\; }}{q} = 2,3},
\end{array}.} \right.
\end{align}

The idea of combining the independent variables before constructing the univariate functions has inspired our improvement of KAN in handling ridge functions.

\subsection{Kolmogorov-Arnold networks}
Kolmogorov-Arnold representation theorem has been interpreted as a feed-forward neural network with
an input layer, one hidden layer, and an output layer~\cite{braun2009constructive,sprecher1996numerical,sprecher1997numerical}. 
Recently,  Liu et al.~\cite{liu2024kan} proposed the Kolmogorov-Arnold network by approximating the univariate functions with splines and generalizing the original representation theorem to arbitrary widths and depths. 
In KAN, the univariate functions in the Kolmogorov-Arnold representation theorem are defined as a linear combination of local B-spline basis functions as follows
\begin{equation}
\phi_{q,m}(x_m) \approx \sum_{i=1} c_{q,m}^i B^i(x_m),
\end{equation}
where \( B^i\) are the local B-spline basis functions. \( c_{q,m}^i \) are learnable coefficients that can be obtained through training.

KAN consists of multiple layers. The set of univariate functions in the $l$-th layer is expressed as
\begin{equation}
\Psi^l = \{ \phi_{s,t}^l \}, \quad s = 1, 2, \ldots, N_{l+1}, \quad t = 1, 2, \ldots, N_{l},
\end{equation}
where the subscripts  \(s\) and \(t\) index the output and input, respectively. \( N_{l+1}\) and \( N_{l}\) represent the number of output and input in the $l$-th layer layer.

The relationship between the input $\mathbf{x}^l = [x_1^l,x_2^l, \ldots]^T$ and output $\mathbf{x}^{l+1} = [x_1^{l+1},x_2^{l+1}, \ldots]^T$ of the \(l\)-th KAN layer is described as
\begin{equation}
{x}_{j}^{l+1} = \sum_{i=1}^{N_l} \phi_{j, i}^l({x}_{i}^l), \quad j = 1, \ldots, N_{l+1}, \quad i = 1, 2, \ldots, N_{l},
\end{equation}
where  \(j\) and \(i\) index the output and input variables, respectively. 

To provide a more intuitive understanding of the KAN neural network, the inputs and outputs of the \(l\)-th layer in matrix form can be denoted as follows
\begin{equation}
\mathbf{x}^{l+1} = \underbrace{\begin{pmatrix}
\phi_{1,1}^l(\cdot) & \phi_{1,2}^l(\cdot) & \cdots & \phi_{1,N_l}^l(\cdot) \\
\phi_{2,1}^l(\cdot) & \phi_{2,2}^l(\cdot) & \cdots & \phi_{2,N_l}^l(\cdot) \\
\vdots & \vdots & \ddots & \vdots \\
\phi_{N_{l+1},1}^l(\cdot) & \phi_{N_{l+1},2}^l(\cdot) & \cdots & \phi_{N_{l+1},N_l}^l(\cdot)
\end{pmatrix}}_{\mathbf{\Phi}^l} \mathbf{x}^l,
\end{equation}
where \(\mathbf{\Phi}^l\) is the function matrix of the \(l\)-th layer.

The deep KAN of $L$ layers with the input $\mathbf{x}$ can then be represented as
\begin{equation}
\text{KAN}(\mathbf{x}) = \mathbf{\Phi}^{L-1} \left\{ \mathbf{\Phi}^{L-2} \left\{ \cdots \left\{ \mathbf{\Phi}^1 \left[ \mathbf{\Phi}^0 (\mathbf{x}) \right] \right\} \right\} \right\}.
\end{equation}

\subsection{Active subspace method}
The active subspace method~\cite{constantine2014active} is a useful technique for identifying the primary directions of variation in a multivariate function according to the function's gradient.
Consider the multivariate continuous function $f(\mathbf{x})$, with $\mathbf{x} = [x_1,x_2, \ldots, x_N]^T$. 
The gradient of function \( f \) can be expressed as
\begin{equation}
    \nabla f = [\frac{{\partial f}}{{\partial {x_1}}},\frac{{\partial f}}{{\partial {x_2}}},\ldots,\frac{{\partial f}}{{\partial {x_{N_f}}}}]^T.
\end{equation}
Let \( \mathbf{W}=[\mathbf{w}_1,\mathbf{w}_2,...,\mathbf{w}_{Nf} \)] be a matrix whose  $N_f$ orthogonal column vectors characterize the direction in which function \( f \) is sensitive to changes.
This means that the projection of $\nabla f$  onto the first to last vector in $\mathbf{W}$  should have the maximum to minimum expected value. We can define the corresponding optimization problem as follows,
\begin{equation}
\begin{split}
	 &\text{maximize} \quad \mathbb{E}[(\nabla f \cdot \mathbf{w})^2], \\
     &\text{subject to} \quad \|\mathbf{w}\| = 1.
\end{split}
\end{equation}
According to the proof of Constantine et al.~\cite{constantine2014active}, this optimization problem can be transformed into solving the eigenvalue problem of the matrix
\begin{equation}
    \mathbf{C} = \mathbb{E}[{( \nabla f)( \nabla f)^T}].
    \label{eq9}
\end{equation}
By performing the following eigenvalue decomposition on $\mathbf{C}$, we obtain
\begin{equation}
    \mathbf{C} =\mathbf{WGW}^T.
\end{equation}
Here, \( \mathbf{W} \) is the matrix of eigenvectors of matrix \( \mathbf{C} \), and \( \mathbf{G} \) is a diagonal matrix with the eigenvalues of \( \mathbf{C} \) on its diagonal. We can identify the primary directions of variation in function \( f \) by solving the eigenvalue problem.

\subsection{Active subspace embedded Kolmogorov-Arnold networks (asKAN)}
\begin{figure}[hbt!]
	\centering    \includegraphics[width=0.7\textwidth]{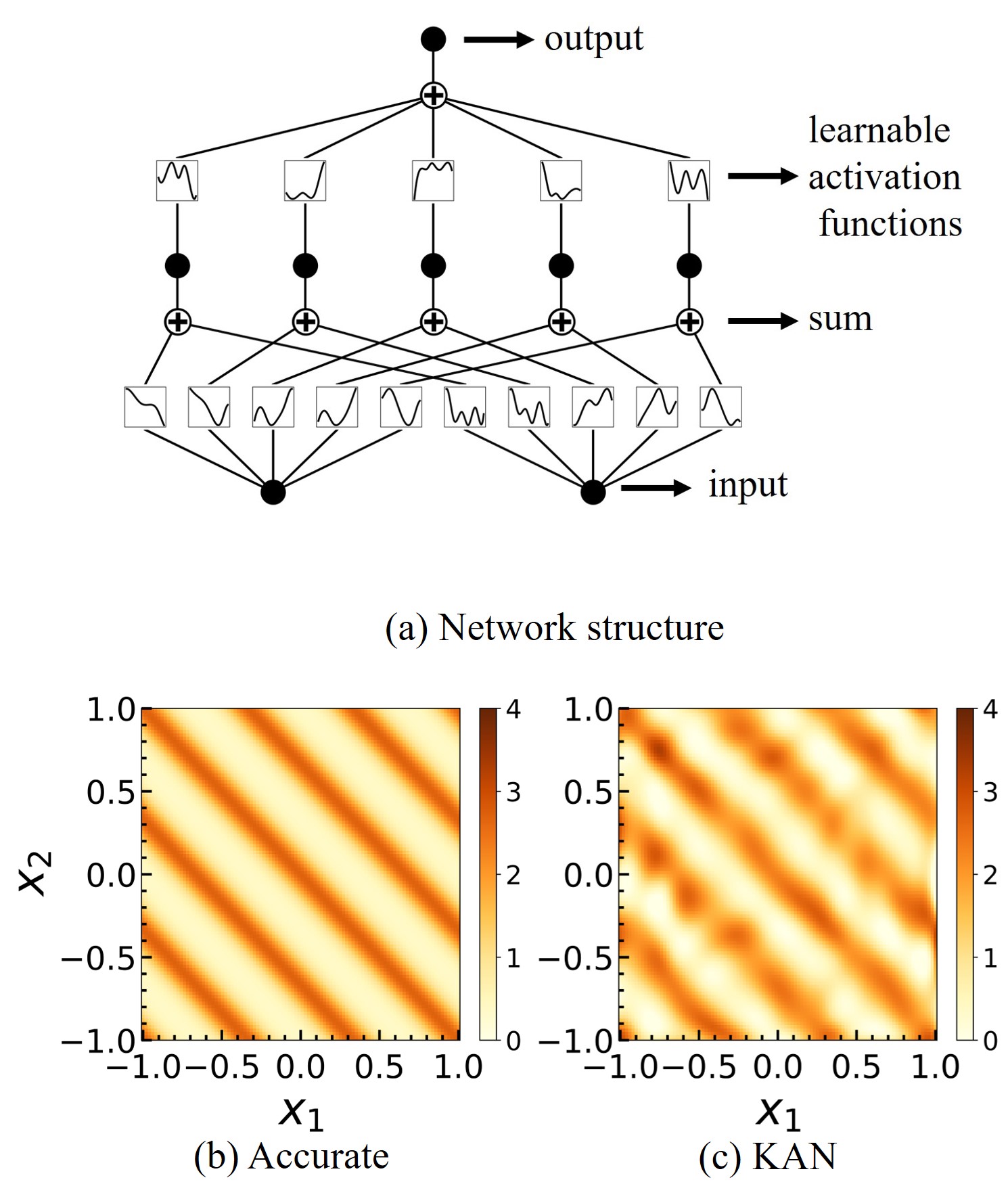}
	\caption{network architecture of KAN and comparison of function values. (a) KAN network architecture; (b) Accurate function values; (c) KAN network fitting results; }
	\label{figa}
\end{figure}

The first layer of KAN is featured by computing the activation function, rather than performing linear composition on the independent variables as MLPs. This distinct feature of KAN leads to inflexibility in handling ridge functions in the form of \( g(x_1, x_2) = f(x_1 + x_2) \). 
For example, the representation of $g(x_1, x_2) = \exp \left( \cos \left( 3\pi (x_1 + x_2) \right) \right)$ using KAN with five nodes in the hidden layer results in spurious oscillations, with a Mean Relative Error (MRE) of 27.4\%, as shown in Figure~\ref{figa}.
The structure of KAN is the same with that used in the work of Liu et al.~\cite{liu2024kan}. The interpolation function is chosen as a third-order spline function, with 5 grid points for each univariate function.

\begin{figure}[hbt!]
	\centering    \includegraphics[width=1\textwidth]{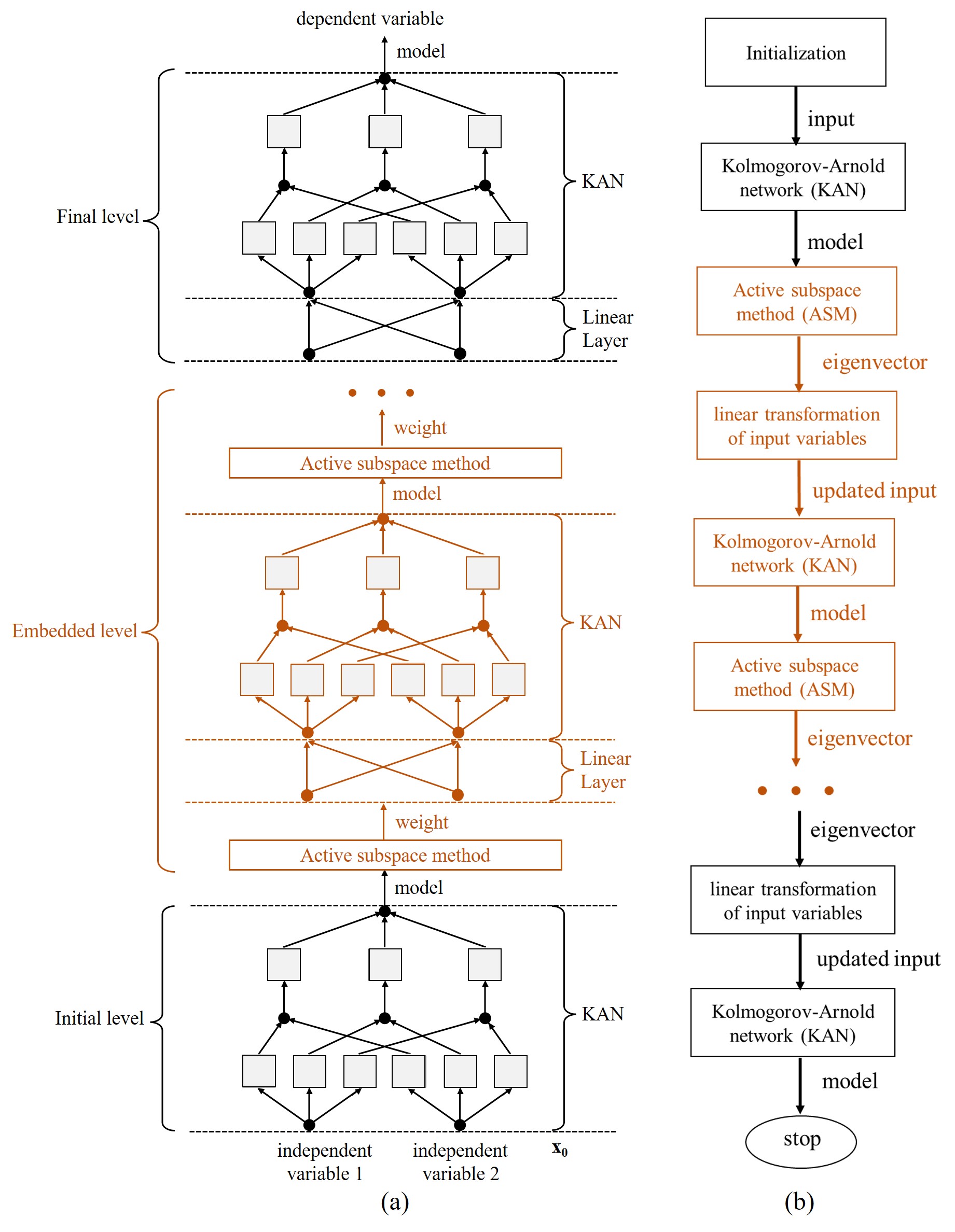}
	\caption{(a) Schematic and (b) flowchart of Active Subspace embedded Kolmogorov-Arnold Network (asKAN)}
	\label{fig0}
\end{figure}

While the accuracy of the representation can be improved by increasing the number of layers or nodes, this work proposes a method without increasing the number of layers or nodes.
we address this issue by performing linear combinations of inputs to identify the primary directions of the multivariate function. Considering that the active subspace method has been widely used for identifying primary directions, we propose to embed the active subspace method into KAN network in a hierarchical framework as follow, 
\begin{equation} 
     {\rm{asKAN}}({\bf{x}}) = {\rm{KAN}}\left\{ {{\bf{W}}_L^T\left\langle {{\rm{KAN}}_L^{}} \right\rangle \left\{ { \cdots \left\{ {{\bf{W}}_1^T\left\langle {{\rm{KAN}}_1^{}} \right\rangle \left[ {{\bf{W}}_0^T\left\langle {{\rm{KAN}}_0^{}} \right\rangle ){\bf{x}}} \right]} \right\}} \right\}} \right\},
\end{equation}
 where $L$ is the number of hierarchy levels and ${{\bf{W}}_i^T\left\langle {{\rm{KAN}}_i^{}} \right\rangle }$ represents the matrix of  eigenvectors obtained at the  $i$-th level. 
The network architecture of asKAN is shown in Fig.~\ref{fig0}(a) and the detailed training flowchart is shown in Fig.~\ref{fig0}(b). The asKAN is constructed in a hierarchical way and implemented in an iterative way. As shown in Fig.~\ref{fig0}, the initial input of asKAN is the raw data \( \mathbf{x}_0 \), which is used to train the initial KAN$_0$:
\begin{equation}
    y = \text{KAN}_0(\mathbf{x}_0).
\end{equation}
Based on the initial model KAN$_0$, we can compute the matrix \(\mathbf{C}\) according to Eq.~(\ref{eq9}). By performing eigenvalue decomposition on \(\mathbf{C}\), we can identify the  matrix of eigenvectors $\mathbf{W}_0$.  Projecting the independent variable \(\mathbf{x}_0\) onto the eigenvectors in the matrix \(\mathbf{W}_0\) results in a new independent variable $\mathbf{x}_1 = \mathbf{W}_0^T \mathbf{x}_0\ $. The new independent variable $\mathbf{x}_1$ 
is then fed into the KAN network to obtain a new model KAN$_1$. We can iteratively repeat this process to improve the  representation. 

	


\section{Validation and discussion}
In this section, we validate the effectiveness of asKAN through function fitting and equation solving. First, we employ asKAN to fit a manually constructed ridge function. Subsequently, we apply asKAN to solve the Poisson equation and compare it with the results of KAN. Finally, we utilize asKAN for the task of sound field reconstruction. 

\subsection{Function fitting}
We use KAN and asKAN to fit a two-dimensional ridge function similar to the sample in Section 2 as follows:
\begin{equation}
    g(x_1, x_2)  = \exp \left( {\cos \left( {3\pi \left( {1.2{x_1} + 0.6{x_2}} \right)} \right)} \right).
    \label{eq12}
\end{equation}
Here \( x_1 \) and \( x_2 \) are the independent variables. We restrict the range of \( x_1 \) and \( x_2 \) to \([-1,1]\). The dataset of this ridge function is divided into a training set and a testing set, each containing 1000 data points. Following the work of~\cite{liu2024kan}, we adopt a network architecture with the shape \{2,5,1\} for training. The interpolation function is chosen as a 3rd-order spline function, with 5 grid points for each univariate function.
\begin{figure}[hbt!]
	\centering    \includegraphics[width=1.0\textwidth]{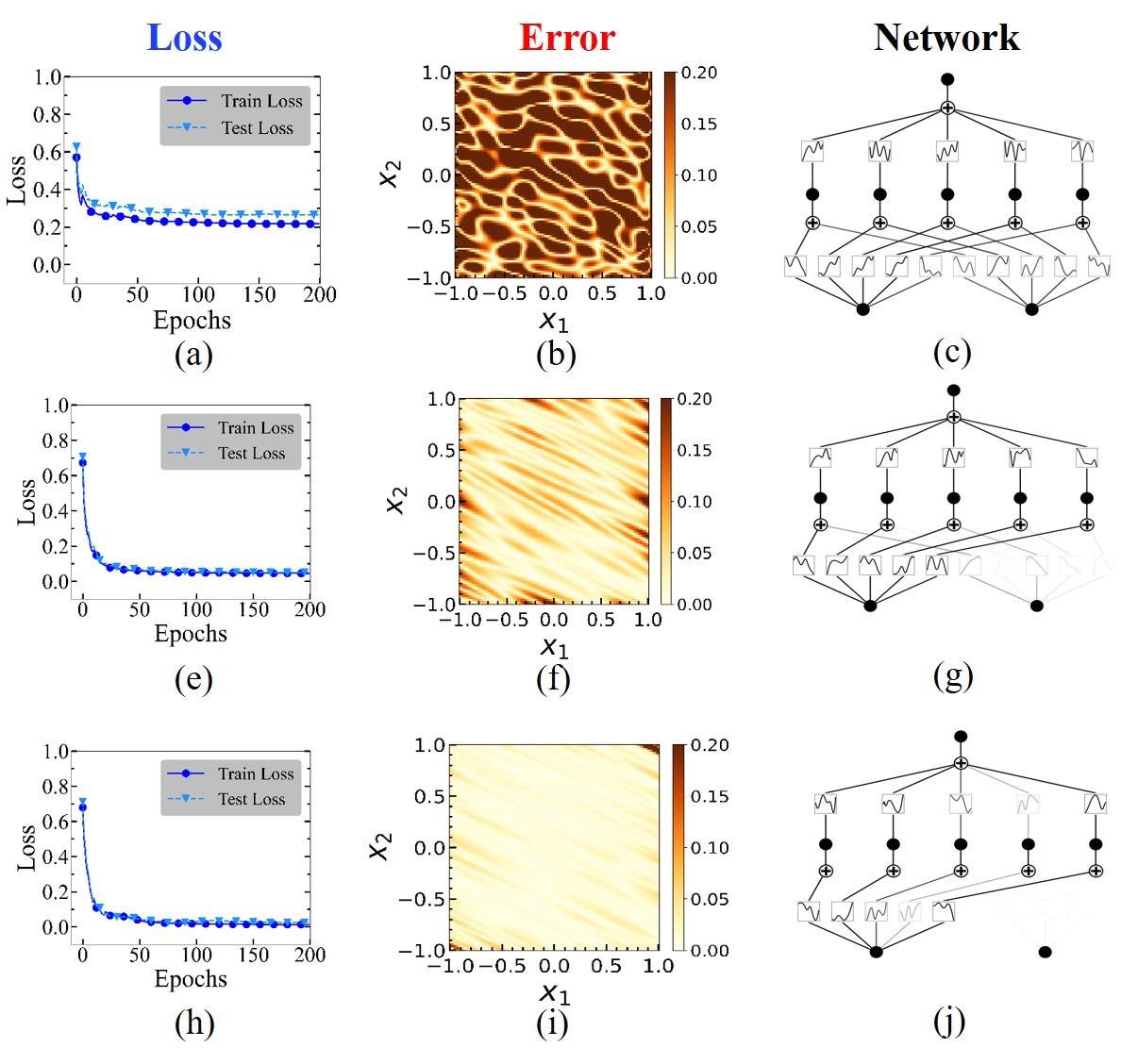}
	\caption{The fitting errors and the obtained networks for the ridge function (Eq.~\ref{eq12}) are presented for the three-level asKAN algorithm. }
	\label{fig2}
\end{figure}

Figure~\ref{fig2} compares the results of the asKAN and KAN, where the asKAN consists of three levels of hierarchy and the first level is the traditional KAN. 
The leftmost column shows the convergence history of training and testing errors, which defined as follows
\begin{equation}
    {\text{Loss} = \mathbb{E}\left[(g_{\text{fit}} - g_{\text{true}})^2\right]}.
\end{equation}
The symbol $\mathbb{E}$ is used to denote the mean value. $g_{\text{fit}}$ and $g_{\text{true}}$ represent the fitted and exact results, respectively. The middle column illustrates the variation of absolute error with the independent variables, and the rightmost column depicts the updated network architecture for the function fitting. As can be seen from Fig.~\ref{fig2}, with the increase of the level of hierarchy in asKAN, the error of the function fitting is significantly reduced. The testing error at convergence decreases by one order of magnitude, from the order of $O(10^{-1})$ to $O(10^{-2})$, and the network architecture becomes increasingly close to that of a univariate function. These results suggest that asKAN can significantly improve the accuracy in fitting the ridge function. 

\subsection{Solving Poisson equation}
The capability of asKAN in solving partial differential equations by considering the Poisson equation with zero Dirichlet boundary data, which is also employed by Liu et al.~\cite{liu2024kan} to test the capability of KAN,
\begin{equation}
\begin{split}
\frac{\partial^2 u}{\partial x_1^2} + \frac{\partial^2 u}{\partial x_2^2} = &\frac{1}{2\pi^2}\sin(\pi x_1) \sin(\pi x_2), \quad (x_1, x_2) \in \Omega = [-1, 1] \times [-1, 1],\\
&u(x_1, x_2) = 0, \quad \forall (x_1, x_2) \in \partial \Omega.
\end{split}
\end{equation}
Here, \(u\) represents the solution of the Poisson equation, while \(x_1\) and \(x_2\) denote the independent variables. The ranges of \(x_1\) and \(x_2\) are both within \([-1,1]\). The Poisson equation in the aforementioned form has an analytical solution as follows
\begin{equation}
    {u_{\text{true}}(x_1, x_2) = \sin(\pi x_1) \sin(\pi x_2)}.
\end{equation}
It is noted that the analytical solution can be turned to  
\begin{equation}
u_{\text{true}}(x_1, x_2) = \frac{1}{2} \left[ \cos(\pi (x_1 - x_2)) - \cos(\pi (x_1 + x_2)) \right],
\label{eq15}
\end{equation}
which is a superposition of two intrinsically low-dimensional functions.
Consistent with Liu et al.'s approach~\cite{liu2024kan}, we set the loss function as the sum of the weighted boundary conditions and the equation residual as follows
\begin{equation}
    {\text{Loss} = \alpha \cdot \mathbb{E}\left[(\nabla^2 u_{\text{pred}} - S)^2] + \mathbb{E}[(u_{\text{pred}} - u_{\text{true}})^2\right]}.
    \label{eq14}
\end{equation}
Here, $\alpha$ is a predetermined weighting coefficient, which is set to $0.01$ in this work. $S=\frac{1}{2\pi^2}\sin(\pi x_1) \sin(\pi x_2) $ is the right-hand-side of the Poisson equation. $u_{\text{pred}}$ represents the predicted solution of the Poisson equation. We utilized a network architecture with the shape \{2,3,1\} for training. The interpolation function was selected as a third-order spline function, with five grid points for each univariate function. 
\begin{figure}[hbt!]
	\centering    \includegraphics[width=0.8\textwidth]{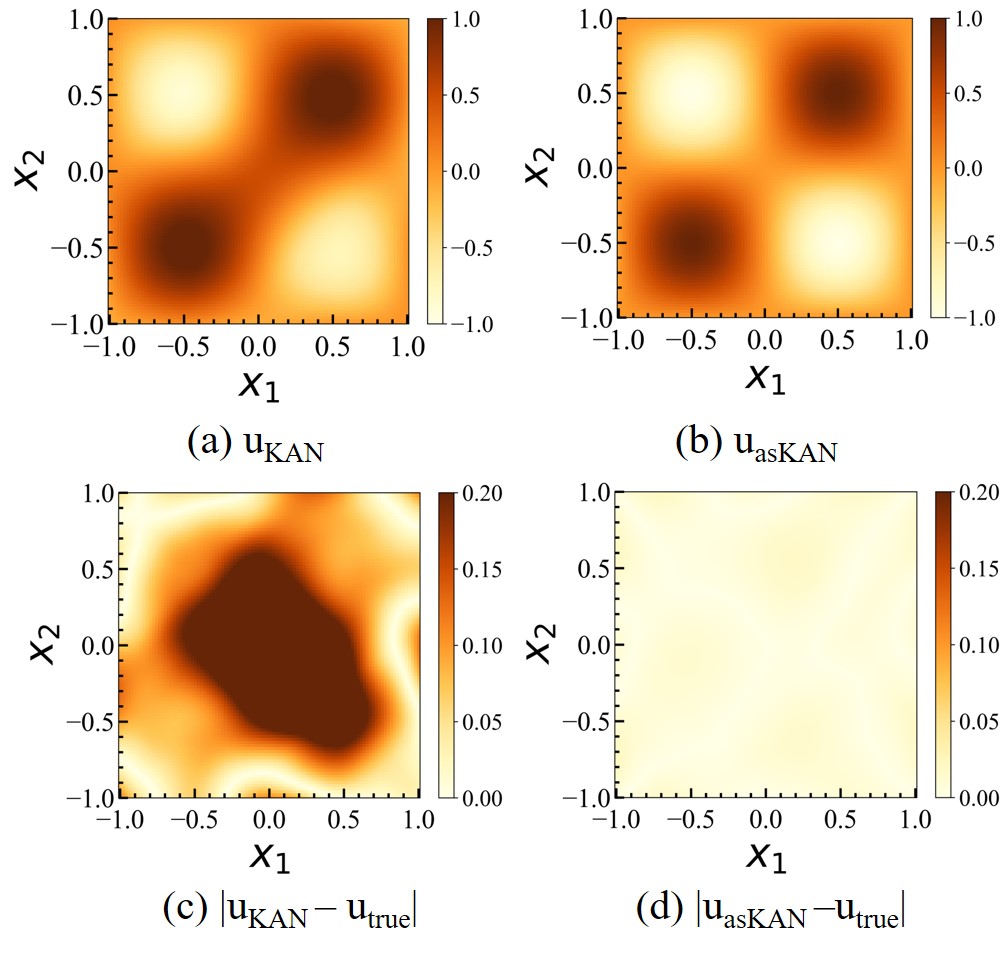}
	\caption{The predicted solution and absolute errors for Poisson equation by using KAN and asKAN. (a) and (b) are the predicted solutions of KAN and asKAN, respectively. (c) and (d) are the absolute errors of KAN and asKAN, respectively. }
	\label{fig3}
\end{figure}

\begin{figure}[hbt!]
	\centering    \includegraphics[width=0.7\textwidth]{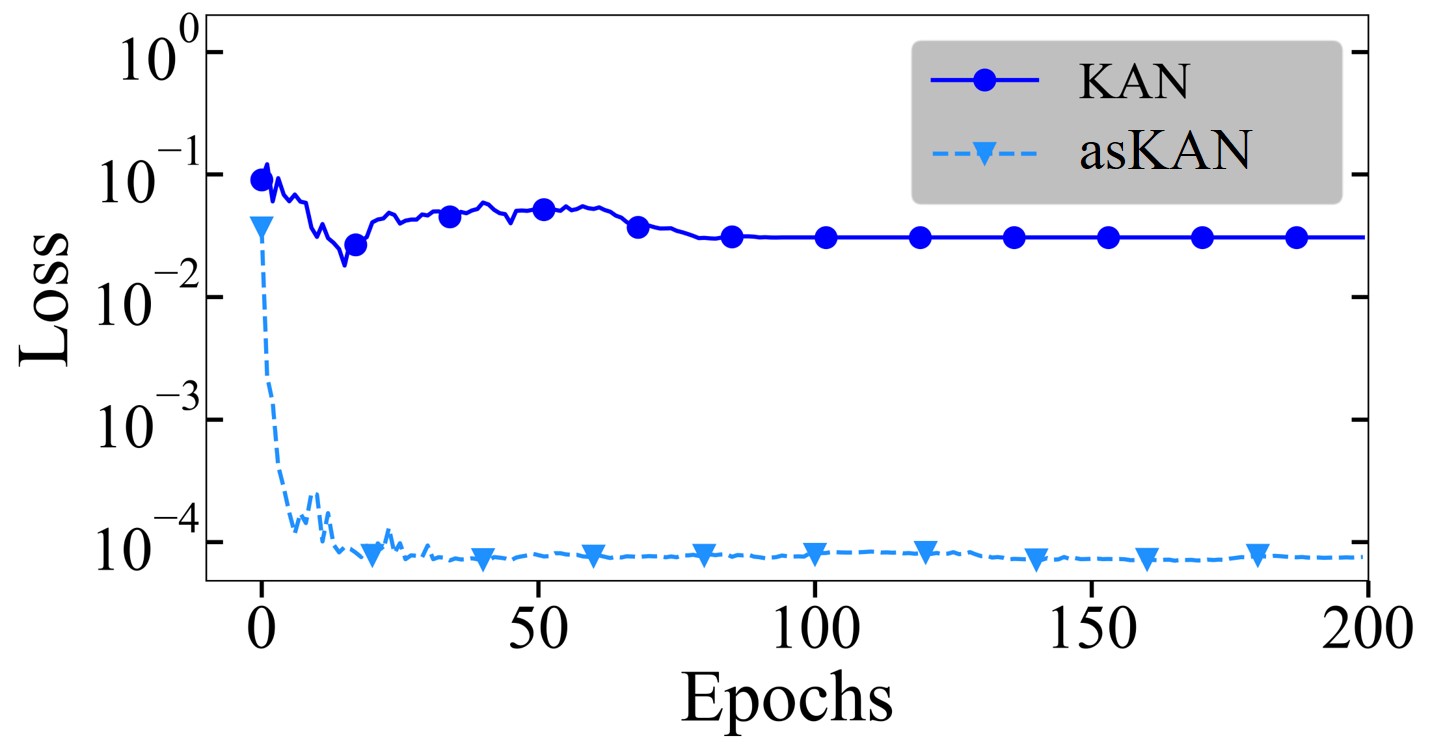}
	\caption{Changes in loss of the KAN and asKAN with epoch for solving Poisson equation.}
	\label{fig4}
\end{figure}

Figure~\ref{fig3} compares the results of KAN and asKAN. Figures~\ref{fig3}(a) and~\ref{fig3}(b) represent the trained function of KAN  and asKAN, respectively. Figures~\ref{fig3}(c) and~\ref{fig3}(d) represent the absolute error between the trained function and the analytical solution \(u_{\text{true}}\). 
The error of asKAN is significantly smaller than that of KAN. Figure~\ref{fig4} compares the convergence history of the loss function for the \(u_{\text{KAN}}\) and \(u_{\text{asKAN}}\). As shown in Fig.~\ref{fig4}, the use of asKAN transforms the loss function of KAN from the order of \(10^{-2}\) to \(10^{-4}\). This result suggests that asKAN can significantly enhance the accuracy in solving the Poisson equation.

\subsection{Scattering sound field reconstruction}

\begin{figure}[hbt!]
	\centering    \includegraphics[width=0.7\textwidth]{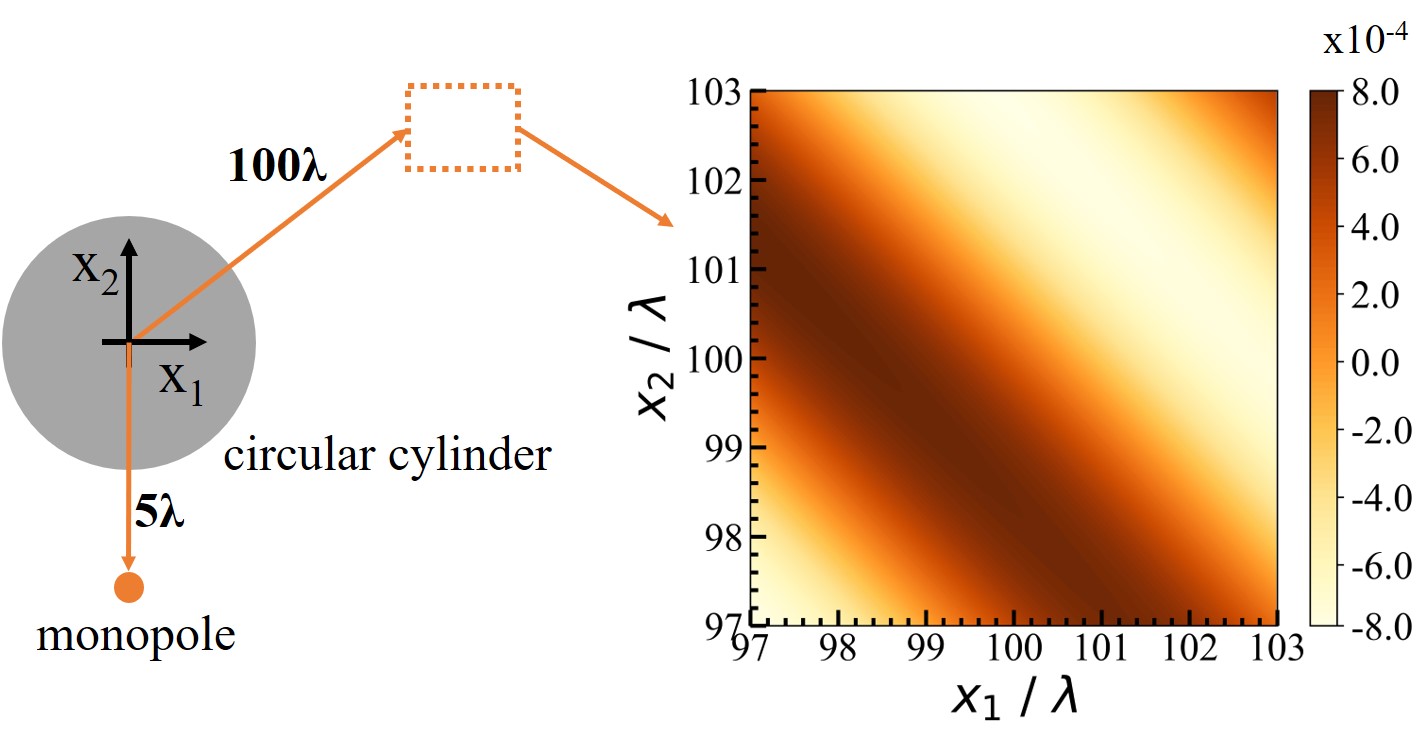}
	\caption{Schematic diagram of the setup for the scattering sound problem}
	\label{fig4p5}
\end{figure}

In this subsection, asKAN is used to reconstruct a sound field based on sparse data. We consider the reconstruction of an acoustic pressure field scattered from a rigid circular cylinder. The speed of sound is defined as $c$. As shown in Fig.~\ref{fig4p5}, a monopole source of frequency $f_s$  is placed at a distance of \(5\lambda\) from a rigid cylinder, where $\lambda$ = $c/f_s$ is the wavelength of acoustic wave. The interaction between the incident sound wave and the cylinder wall generates scattering noise, which satisfies the wave equation
\begin{equation}
    \frac{\partial^2 p}{\partial t^2} = c^2 \left( \frac{\partial^2 p}{\partial x_1^2} + \frac{\partial^2 p}{\partial x_2^2} \right).
\end{equation}
Here, \(p\) represents the fluctuating sound pressure, \(x_1\) and \(x_2\) denote the spatial coordinates. We sample the sound pressure in a square region with a side length of \(6\lambda\) at a distance of \(100\lambda\) from the circular cylinder. Within this square region, we uniformly distribute 100 virtual sensors (points) at a sampling frequency of \(20f_s\), resulting \(100 \times 200 = 20,\!000\) spatiotemporal sampling points.

We divide the sampling data into a training set containing 1000 points and a testing set with 19000 points to evaluate the sound field reconstruction capability. The loss function is defined as the sum of the differences in acoustic pressure at the sampling points and the residuals of the wave equation
\begin{equation}
\begin{split}
    \text{Loss} =\mathbb{E}\left[ \left| p_{\text{predicted}}(x_{1},x_{2}, t) - p_{\text{measured}}(x_{1},x_{2}, t) \right|^2\right] + \\
	\alpha \mathbb{E}\left[ \left| \frac{\partial^2 p}{\partial t^2} - c^2 \left( \frac{\partial^2 p}{\partial x_1^2} + \frac{\partial^2 p}{\partial x_2^2} \right) \right|^2\right],  
\end{split}
\end{equation}
where \(\alpha\) is defined as 1 in this case. We use a network architecture with the shape \{3,5,1\} for training. The active function is approximated as a third-order spline function, with five grid points for each univariate function.

\begin{figure}[hbt!]
	\centering    \includegraphics[width=0.8\textwidth]{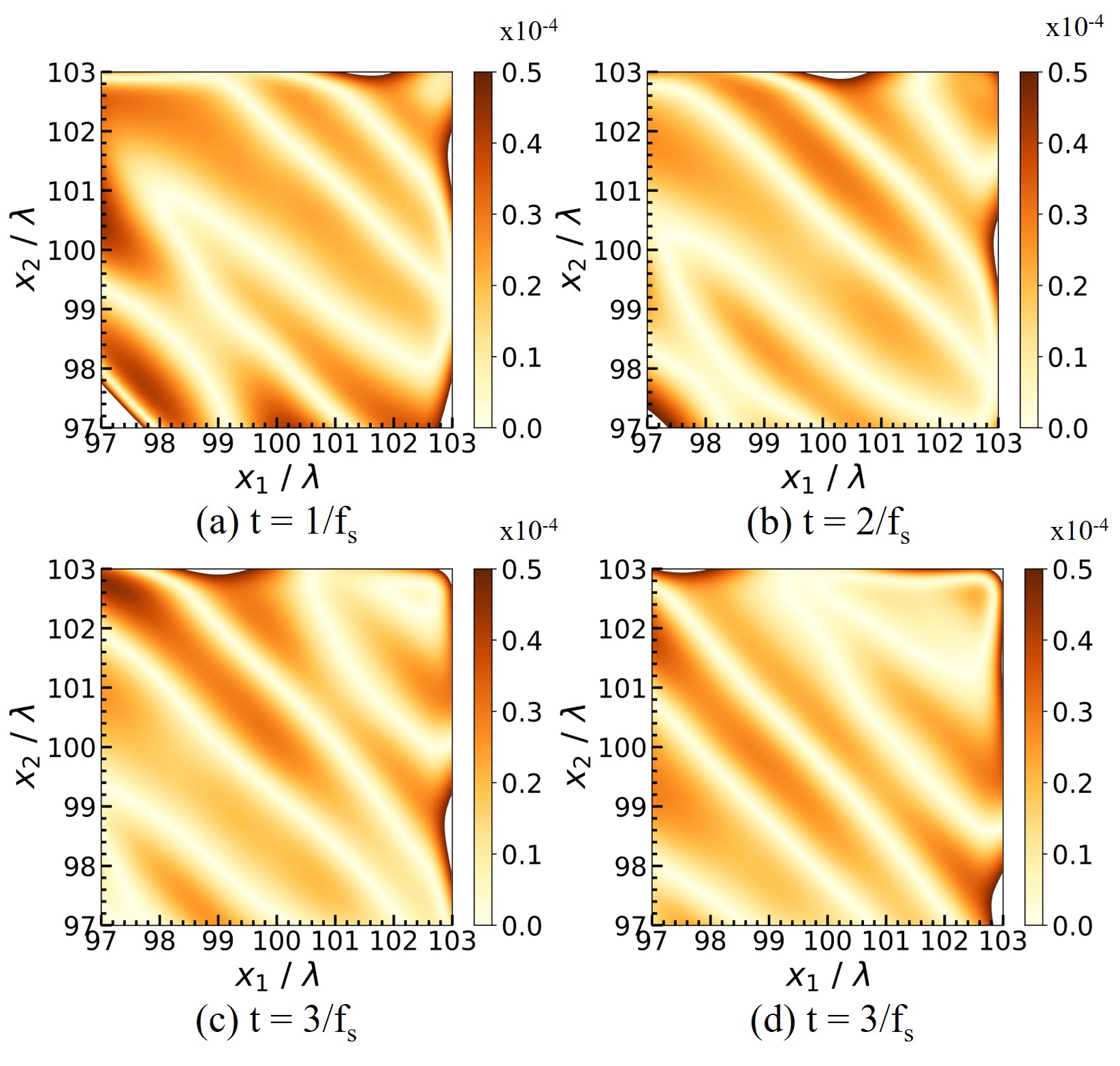}
	\caption{Absolute error for the acoustic pressure predicted by using KAN at different moments}
	\label{fig6}
\end{figure}

\begin{figure}[hbt!]
	\centering    \includegraphics[width=0.8\textwidth]{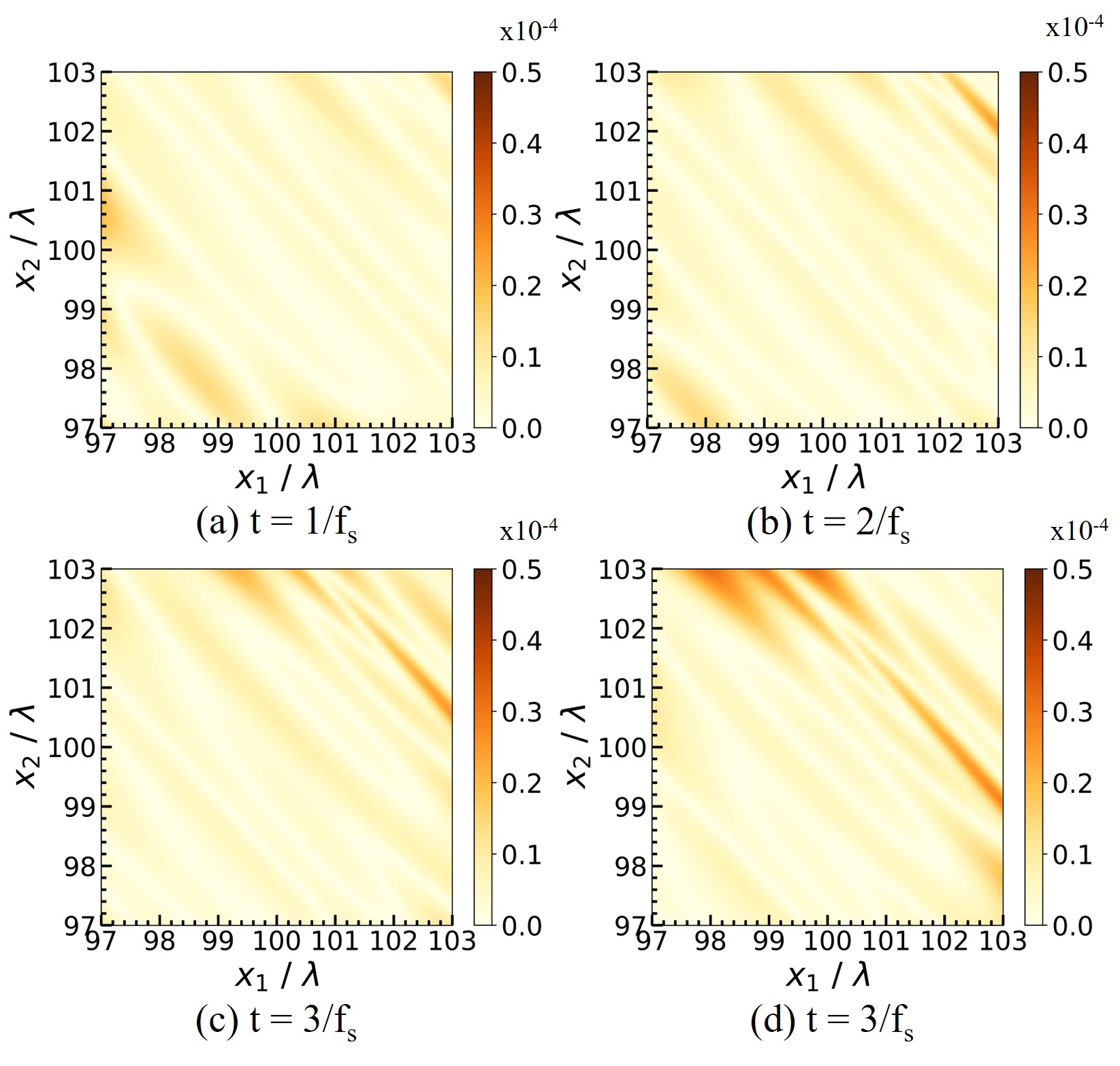}
	\caption{Absolute error for the acoustic pressure predicted by using asKAN at different moments}
	\label{fig7}
\end{figure}

\begin{figure}[hbt!]
	\centering    \includegraphics[width=0.8\textwidth]{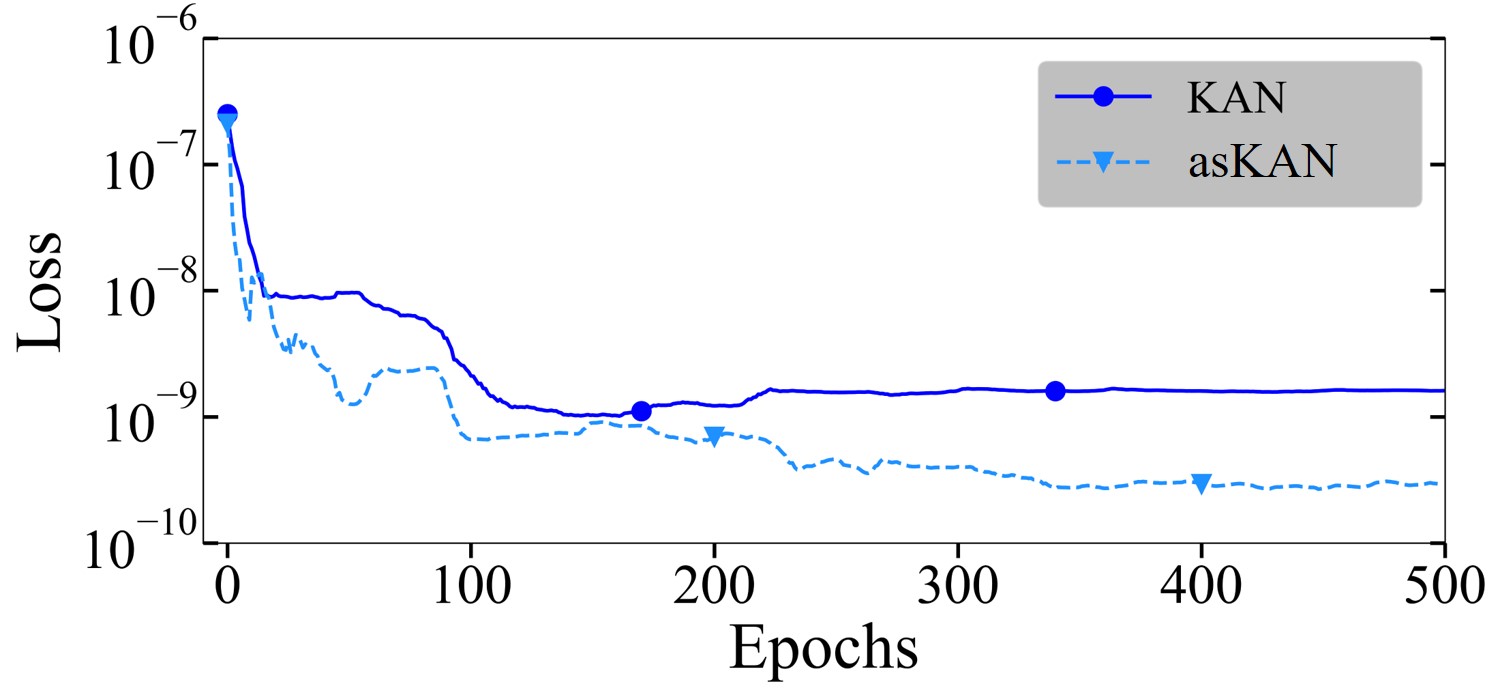}
	\caption{Changes in loss with epoch for the KAN and asKAN for the sound field reconstruction}
	\label{fig8}
\end{figure}

Figures~\ref{fig6} and~\ref{fig7}  show the error of predicted acoustic pressure at different moments by using KAN and asKAN, respectively.  It is shown that the error of asKAN is significantly smaller than that of KAN. Figure~\ref{fig8} compares the convergence history of the loss function for the predicted acoustic pressure. As shown in Fig.~\ref{fig8}, the use of asKAN transforms the loss function of KAN from the order of \(10^{-9}\) to \(10^{-10}\). This result suggests that asKAN can significantly enhance the accuracy of reconstructing the sound field.

\section{Conclusion}

The Kolmogorov-Arnold network (KAN) is a promising neural network that has advantages in small-scale  AI + Science tasks. However, it suffers from inflexibility in representing the ridge functions.
In this work, the inflexibility of KAN is investigated from the feature of Kolmogorov-Arnold representation theorem, which starts from the construction of univariate function instead of combination of the independent variables. 
We found that the introduction of linear combination of independent variables can significantly simplify the representation of ridge functions.
We proposed active subspace embedded KAN (asKAN) to circumvent the inflexibility. The proposed asKAN organizes KANs in a hierarchycal framwork, where the active subspace method is embedded between the neighbouring levels. 
The active subspace method is used to detect the primary directions of the ridge functions by estimating their gradients based on KAN from a previous level. 
Then the input for the next level is improved by projecting the independent variables onto these primary directions.
The proposed asKAN is implemented in an iterative way, which results in a compact low-dimensional neural network.
We validated the proposed asKAN through function fitting, solving the Poisson equation, and reconstructing sound field. Compared with KAN, asKAN significantly reduces the error without increasing the number of neurons. 
The results suggest that asKAN provides valuable assistance in fitting and solving equations in the form of ridge functions.


\section*{Acknowledgements}
This work is supported by the NSFC Basic Science Center Program for “Multiscale Problems in Nonlinear Mechanics” (No.11988102), the National Natural Science Foundation of China (Nos.92252203 and 12102439), the Chinese Academy of Sciences Project for Young Scientists in Basic Research (YSBR-087), and the Strategic Priority Research Program of Chinese Academy of Sciences (XDB0620102).

\bibliography{mybibfile}

\end{document}